\documentclass[aps,
		prd,
		reprint,
		twocolumn,
		superscriptaddress,
		shortbibliography,
		footinbib,
		floatfix
		]{revtex4-1}

\usepackage{amsmath,amssymb,amsfonts}

\usepackage{natbib}
\usepackage[T1]{fontenc}
\usepackage[latin1]{inputenc}

\usepackage[dvipsnames]{xcolor}

\usepackage{hyperref}
\hypersetup{colorlinks=true,citecolor=Cyan,urlcolor=Red}

\usepackage{mathrsfs}
\usepackage{bbm}
\usepackage{slashed}
\DeclareSymbolFont{rsfs}{U}{rsfs}{m}{n}
\DeclareSymbolFontAlphabet{\mathrsfs}{rsfs}
\usepackage[mathscr]{eucal}	
\usepackage[normalem]{ulem}
\usepackage{verbatim}

\usepackage{epsfig}
\usepackage{graphicx}
\usepackage{tikz}
\usetikzlibrary{positioning}

\usepackage{booktabs}

\newcommand{\trm}[1]{\textrm{#1}}

\newcommand{\eqnref}[1]{(\ref{#1})}

\newcommand{\lcfap}{LCFA\texttt{+}~}

\let\vec\boldsymbol

\newcommand{\be}{\begin{equation}}
\newcommand{\ee}{\end{equation}}
\newcommand{\bi}{\begin{itemize}}
\newcommand{\ei}{\end{itemize}}
\newcommand{\bea}{\begin{eqnarray}}
\newcommand{\eea}{\end{eqnarray}}
\newcommand{\ud}{\mathrm{d}}		

\newcommand{\LCperp}{{\scriptscriptstyle \perp}}

\begin{document}

\title{An extended locally constant field approximation for nonlinear Compton scattering}
\author{A. Ilderton}
\affiliation{Centre for Mathematical Sciences, University of Plymouth, Plymouth, PL4 8AA, UK}
\author{B. King}
\affiliation{Centre for Mathematical Sciences, University of Plymouth, Plymouth, PL4 8AA, UK}
\author{D. Seipt}
\affiliation{Physics Department, Lancaster University, Bailrigg, Lancaster LA1 4YW, UK}
\affiliation{The Cockcroft Institute, Daresbury Laboratory, WA4 4AD, UK}
\affiliation{Center for Ultrafast Optical Science, University of Michigan, Ann Arbor, Michigan 48109, USA}
\begin{abstract}	
The locally constant field approximation (LCFA) has to date underpinned the numerical simulation of quantum processes in laser-plasma physics and astrophysics, but its validity has recently been questioned in the parameter regime of current laser experiments. While improvements are needed, literature corrections to the LCFA show inherent problems. Using nonlinear Compton scattering in laser fields to illustrate, we show here how to overcome the problems in LCFA corrections. We derive an ``\lcfap'' which, comparing with the full QED result, shows an improvement over the LCFA across the whole photon emission spectrum. We also demonstrate an implementation of our results in the type of numerical code used to design and analyse intense laser experiments.
\end{abstract}
%

\maketitle
Strong electromagnetic fields are found in intense laser-matter interactions, around astrophysical objects such as magnetars, and in the collision point of particle colliders.  The coupling between particles and a strong field is, by definition, larger than unity and so must be accounted for \textit{non}-perturbatively.  This may be achieved, in the calculation of quantum processes, by employing the Furry expansion of QED scattering amplitudes~\cite{furry51}. Analytically, however, such calculations are limited to simple field models; lasers, for example, are almost universally modelled as plane waves~\cite{volkov35,ritus85,Seipt:2017ckc}. Within this model, calculations involving even a single seed electron are challenging, while experiments typically employ bunches of the order of $10^8$ electrons and laser pulses which are tightly focussed in space, i.e.~far from plane wave. In order to bridge the gap between theory and experiment, particle-in-cell (PIC) simulations are used, in which quantum probabilities are calculated using Monte-Carlo event generators, for a review see~\cite{Gonoskov:2014mda}.  A key ingredient is the locally constant field approximation (LCFA)~\cite{nikishov64,baier94}, which assumes that strong fields can be regarded as ``instantaneously constant'' over the timescales of QED processes. The LCFA allows known scattering amplitudes in constant crossed fields (the zero-frequency limit of plane waves) to be adapted to \textit{arbitrary} fields in simulations, thus aiding experimental programmes.

However, the LCFA's region of validity is limited. Consider nonlinear Compton scattering (NLC), that is, photon emission from an electron in a strong laser field~\cite{nikishov64,harvey09,Boca:2009zz,heinzl10b,seipt11,mackenroth11,Seipt2018}. The LCFA for this process fails in some parts of the emitted photon spectrum~\cite{harvey15}, fails to capture interference effects~\cite{harvey15,dinu16} and, critically, its applicability in interpreting experimental results~\cite{sarri14,cole18,sarri17}  has recently been called into question~\cite{sarri17,meuren17}.  Literature approaches to improving the LCFA are typically based on adding to it corrections in the form of a gradient expansion (of a QED result)~\cite{baier-crystals-86,baier-ultra-89,baier-mono-81}. However, comparisons with the LCFA and QED are lacking. An examination of the corrections (below) reveals that they can give large and unphysical contributions, rather than the expected small corrections which improve on the accuracy of the approximation. These results are not suitable for improving numerics.

In this paper, using NLC as the context, we identify the origin of the problems with the LCFA corrections and, crucially, find a method to resolve them. From this, we derive an improved photon emission rate which reproduces QED results better than the LCFA. We demonstrate its use in (single particle) numerical simulations, and provide a prescription for extending the results to the PIC simulation of particles interacting with realistic focussed pulses.

To begin, consider an electron of initial momentum $p_\mu$ colliding with a plane wave travelling in the $k_\mu$ direction, i.e.~depending on phase $k.x$, of peak intensity parameter $a_0$, central frequency $\omega =k_0$, and arbitrary temporal profile. Define the invariant energy parameter $b=k.p/m^2$, for $m$ the electron mass. The electron emits a photon of momentum $k'_\mu$, which has a ``light-front momentum fraction'' $s=k.k'/k.p$. The NLC probability may be written as an integral over $s$ and two phases, $\varphi$ and $\theta$, the latter of which parametrises interference effects, see~\cite{dinu16} and the appendix for details.

The LCFA is supposed to hold when $a_0\gg 1$ and the electromagnetic field invariants scaled by the Schwinger field are $\ll 1$ and the quantum nonlinearity parameter $\chi:=a_0 b$~\cite{ritus85}.  The NLC probability is then approximated by a phase ($\varphi$) integral over the constant crossed field result but with the field strength replaced by the local field strength depending on $\varphi$. Now, constant field or LCFA results in the literature are obtained by taking a \textit{leading order asymptotic limit} of the full QED results, see~\cite{nikishov64,baier-mono-81,ritus85} for examples.  Therefore the LCFA can be written as the leading term of some asymptotic expansion of the QED result, and corrections to this may, as is standard for asymptotic series, give a better or worse approximation than the leading order (LCFA) term. 
(We will see examples of this below.) Starting from large $a_0$, we show in the appendix that when the parameter $\zeta := s a_0^2(\varphi)/8b(1-s)$~\cite{meuren17} is also large, the asymptotic expansion in $a_0$ can be mapped to a perturbative (or derivative) expansion of the QED result in (small) inverse powers of $\zeta$~\cite{bender78}. In this situation corrections to the LCFA are well behaved and under control. Such corrections have previously been written down~\cite{baier-mono-81,baier-ultra-89}, but those results are not used by the community, and, to the best of our knowledge, have not been implemented in numerical simulations. One reason for this, which has not previously been discussed, is that for $s\to 0$ or $a_0(\varphi)\to 0$ the ``corrections'' become very large, rather than small. This is precisely when the accuracy of the derivative expansion breaks down. The solution we propose is straightforward: in keeping with asymptotic methods, we include derivative corrections when the change is small compared to the LCFA, but not when the change is large. We will see that this strategy makes physical sense and better approximates QED.

\begin{figure}[t!]
\includegraphics[width=\columnwidth]{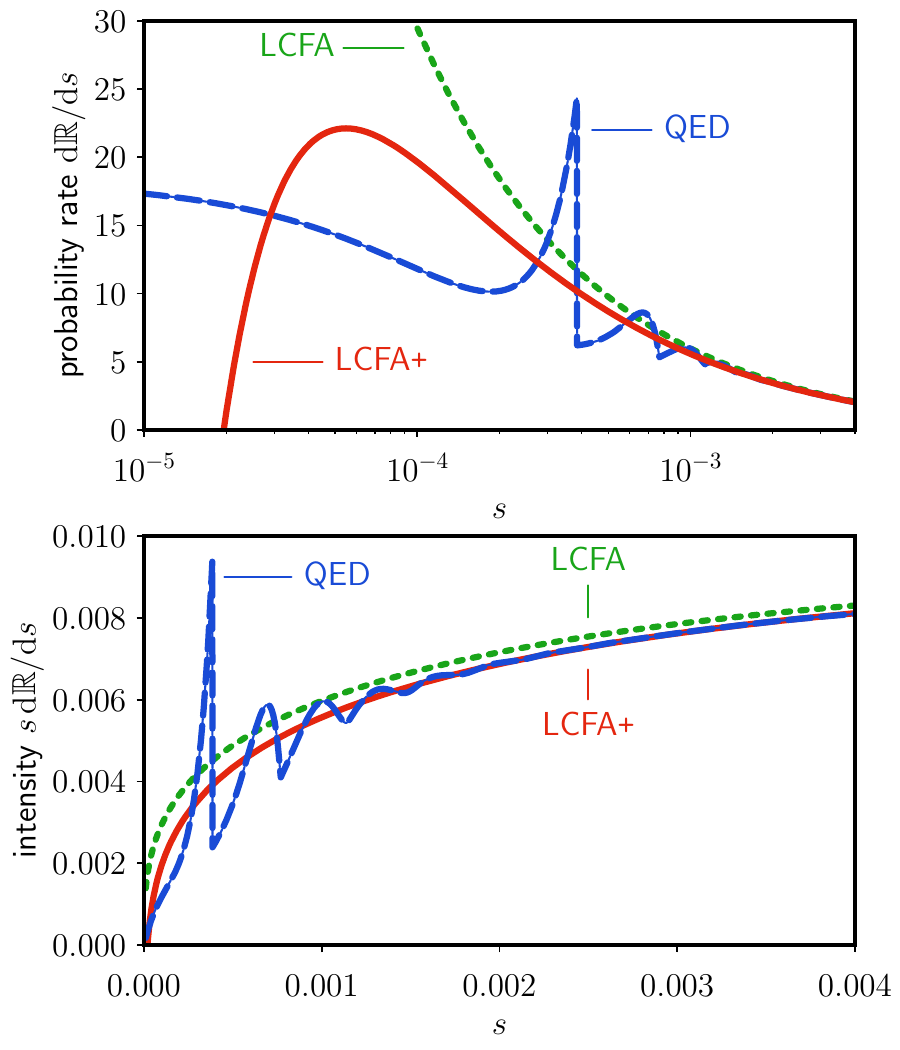}
\caption{\label{FIG:MONO} Upper: the emitted photon spectrum from an electron, $\gamma=1250$, colliding head-on with a monochromatic field of $a_0=5$ and optical frequency $\omega = 1$ eV. Lower:  the emitted (lightfront) energy density, shown at small $s$. The LCFA overestimates the QED result across the spectrum, and is blind to the harmonic structure. The \lcfap follows the QED curves much more closely, and cuts through the harmonics such that the total (integrated) emitted energy agrees better with QED.
}
\end{figure}

We now write down the standard LCFA and its first correction, for arbitrary plane wave fields. Let the two components of the plane wave electric field, made dimensionless, be $\varepsilon_j(\varphi) := e E_j(\varphi)/m\omega = a_0 h_j(\varphi)$ where $a_0$ is the peak absolute value and $h_j$ is a profile function. From this we define the \textit{local} $\chi$-factor of the electron, and local $a_0$, by $\chi_e(\varphi):= \sqrt{\varepsilon_j (\varphi)\varepsilon_j(\varphi)}\,b \equiv a_0(\varphi)b$. The analogous nonlinearity parameter for the emitted photon is $\chi_\gamma(\varphi) = a_0(\varphi)s b$. Define also $z$ and $g$ by
\be\label{eqn:zg} 
		z(\varphi) := \bigg(\frac{1}{\chi_e(\varphi)}\frac{s}{1-s}\bigg)^{\tfrac{2}{3}} \;,\  g(\varphi):= \frac{2}{z(\varphi)}+\chi_\gamma(\varphi)\sqrt{z(\varphi)} \;.
\ee
The LCFA to the probability of NLC is then
%
%
\begin{eqnarray}
\mathbb{P}_\text{LCFA}(\varphi) &=&
	 -\frac{\alpha}{b} \int\!\ud\varphi\! \int\limits_0^1\! \ud s \bigg\{ \text{Ai}_1[z(\varphi)] + g(\varphi)\text{Ai}'[z(\varphi)] \bigg\} \nonumber\\
	 	&=:& \int\!\ud\varphi\ \mathbb{R}_\text{LCFA}(\varphi) \;.\label{Dave-4}
\end{eqnarray}
($\alpha\approx1/137$ is the fine-structure constant.) For a constant crossed field this result is exact, $\mathbb{R}_\text{LCFA} \equiv \mathbb{R}_\text{CCF}$, thus (\ref{Dave-4}) is indeed the LCFA. It depends only on local $\chi_e(\varphi)$ (aside from the flux prefactor $1/b$~\cite{Ilderton:2012qe}). This locality is what allows for the identification of $\mathbb{R}_\text{LCFA}$ as a \emph{photon emission rate}. The first corrections to the LCFA depend explicitly not only on the local value of the electromagnetic fields but also on their derivatives, through the two dimensionless combinations
\be\begin{split}\label{F-def}
	\frac{1}{a_0^2}\mathcal{F}_1 &= \frac{3 \varepsilon_j \varepsilon_j''+\varepsilon_j'\varepsilon_j'}{45\, a_0(\varphi)^4}, \quad
	\frac{1}{a_0^2}\mathcal{F}_2 = \frac{3 \varepsilon_j \varepsilon_j''-4\varepsilon_j'\varepsilon_j'}{45\, a_0(\varphi)^4} .
\end{split}
\ee
Otherwise, the form of the corrections is very similar to that of the LCFA itself; explicitly,
\be\label{delta-LCFA}\begin{split}
	\delta\mathbb{R}(\varphi) = \frac{1}{a_0^2}\frac{\alpha}{b}  \int\limits_0^1\!\ud s\ \mathcal{F}_2(\varphi)  g(\varphi)\bigg(\frac{\text{Ai}(z)}{z}+ \text{Ai}'(z)\bigg) \\
	- \mathcal{F}_1(\varphi)  \bigg(g(\varphi) - \frac{1}{z}\bigg)\bigg(z^2\text{Ai}(z)+2 \text{Ai}'(z)\bigg) \;.
\end{split}
\ee
Let us illustrate the problems of naively using the LCFA and its corrections by considering a monochromatic, circularly polarised field. In this case corrections equivalent to (\ref{delta-LCFA}) were written down in~\cite[Eq.~(4.16)]{baier-mono-81}, but no comparison with the LCFA was made, nor were problems with the corrections highlighted or resolved. The generic forms of the QED photon spectrum~\cite{nikishov64,harvey09}, the LCFA, and the corrected LCFA including \eqref{delta-LCFA}, or ``\lcfap'' are shown in Fig.~\ref{FIG:MONO}. We have $\mathcal{F}_1(\varphi)=-2/45$ and $\mathcal{F}_2(\varphi) = -7/45$, constants.  First, neither the LCFA nor the \lcfap recover harmonic structure at low $s$~\cite{harvey15,meuren17}. The reason is that this structure is generated by contributions from large $\theta$~\cite{harvey15}, while (see the appendix) the LCFA is explicitly tied to a \textit{small} $\theta$ expansion~\cite{Khokonov:PRL2002,harvey15,meuren17,Blackburn:2018sfn}. The second problem of the LCFA is that it over-estimates the QED result at larger~$s$. We can clearly see, though, that the \lcfap solves this problem of over-estimation, agreeing much more closely with the QED result. It cuts the `middle' of the harmonic structure and so, as we have verified, integrated observables such as the total emitted energy agree much more closely with QED than those of the LCFA. This improvement holds down to small~$s$ where, from the introductory discussion, we expect things to break down. Here the \lcfap rate becomes infinitely negative, as opposed to infinitely positive in the LCFA, but a rate corresponding to probability per unit time clearly cannot be negative. This problem comes from the correction term $g \text{Ai}/z \sim \text{Ai}/z^2$ in (\ref{delta-LCFA}). This diverges like $s^{-4/3}$ at small~$s$, which is worse than the LCFA, where the singularity goes like $s^{-2/3}$ and is integrable. The technical reason for these behaviours is that expanding the QED result in the parameter of~\cite{meuren17} requires Taylor expanding Kibble's effective mass~\cite{Kibble:NPB1975} in powers of $\theta$. The mass asymptotes to a finite value as $\theta \to \infty$~\cite{harvey12}, but any order of the expansion naturally gives a power law dependence, with the approximated mass diverging to infinity more rapidly the higher the order of expansion taken \footnote{For a finite pulse it approaches the electron rest mass $m$, while for an infinite plane wave the limiting value is the intensity-dependent effective mass $m_\star$ \cite{harvey12}.}. (See Fig.~\ref{FIG:M2-EXPANSION} in the appendix.) Because it is large~$\theta$ which determines the small~$s$ behaviour of the photon spectrum~\cite{harvey15}, poorly approximating the former introduces errors in the latter.

The above illustrates that naively applying corrections to the LCFA gives some improvements, but that problems remain. Further problems are revealed by looking at the more physical case of pulsed fields. Then (in contrast to the monochromatic case) the $\mathcal{F}_j(\varphi)$ will in general blow up, independent of $s$, when the field strength goes to zero, $a_0(\varphi)\to0$, as it does both outside the pulse and also whenever the field oscillates.  Numerical investigation shows that this is where the apparently small corrections become large: the corrected rate (\ref{delta-LCFA}) exhibits very large peaks which exceed the LCFA result, and which do not appear in the full QED result. (The Airy functions go to zero exponentially faster in the same limit, so there is no divergence, but we can still have $\mathcal{F}_j$ large while the Airy functions remain small, leading to the large peaks.)

Physically, we expect only low emission from regions of very small $a_0(\varphi)$, and hence negligible contributions to the total probability. Observe that, despite the nature of its asymptotic series, the leading order LCFA gets this behaviour right, returning zero at low intensity; when $\chi_e \sim a_0(\varphi) \to 0$ the Airy functions go to zero exponentially quickly (and there are no prefactors), killing low-intensity contributions. In this way the LCFA ``self-regulates'', removing contributions from small~$a_0(\varphi)$ which we expect to be small. Thus, in this situation, the leading order asymptotic result (the LCFA) is enough. Thus we do not include the correction $\delta \mathbb{R}$ at low intensity. Practically, this just means `filtering' the correction (\ref{delta-LCFA}) by multiplying  it by $\mathbf{\Theta}_{i} := \Theta(a(\varphi)-c)$ for some positive constant~$c$  to be determined, below.


Returning now to the low-$s$ behaviour, we also need to make sure that the rate stays positive. (This is not fulfilled even by the full QED ``rate'' $\ud\mathbb{P}/\ud\varphi$ because of quantum interference effects, which frequently give negative contributions~\cite{king15b}. However, our interest is not in directly approximating QED observables, which are readily calculable by other means~\cite{dinu13a,Seipt:JPP2016}, but in generating an improved rate suitable for eventual implementation in Monte Carlo (MC) codes.) Because of its singular behaviour at low~$s$, the magnitude of the LCFA correction exceeds that of the LCFA below some small $s$, and the corrected rate becomes negative. (This assumes $\mathcal F_2(\varphi) <0$, which seems to be the generic case; fringe cases are discussed in the appendix.) Hence we again exclude corrections to the LCFA, as for low intensity, above. Unlike for low intensity, though, the standard LCFA does not gives a physically sensible result, diverging at low-$s$ instead of going to a constant~\cite{meuren17,Blackburn:2018sfn}. There is therefore still a need to fix the low-$s$ behaviour of the LCFA itself.

\begin{figure}[t!]
	\begin{center}
	\includegraphics[width=0.99\columnwidth]{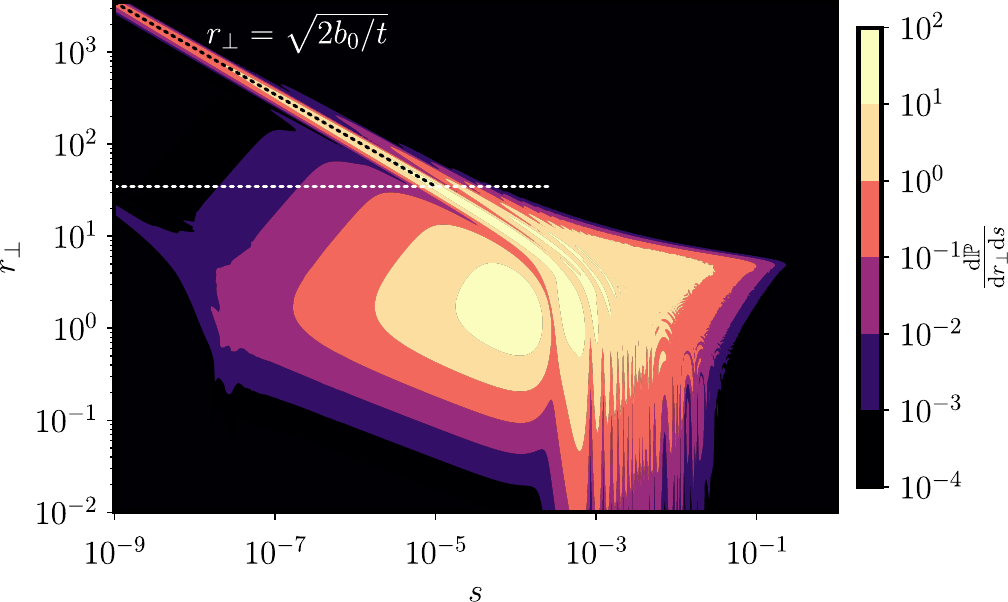}
	\end{center}
	\caption{
		Double differential QED probability showing that low-$s$ corresponds to large $r_\perp$, i.e.~large angles. Parameters: $\gamma=1000$, $a_0 = 5$, short pulse envelope $g=\cos^2(\varphi/8) \Theta(4\pi - |\varphi|)$, circular polarisation. The finite value of the spectrum as $s\to0$ comes from the first harmonic, for large angles, which can be described by the usual Klein-Nishina formula for linear Compton scattering. The horizontal dashed line shows the angular cutoff $r_\perp = 7 a_0$. 
	\label{fig:QED-angle}	 
	 }
\end{figure}

Consider Fig.~\ref{fig:QED-angle}, which shows the double-differential QED spectrum $\ud^{2} \mathbb P /\ud s\, \ud r_\perp$ as a function of $s$ and the dimensionless transverse photon momentum $\vec r_\perp = \vec k_\perp'/ms$; this has magnitude $r_\perp = \frac{p^+}{m}\tan \frac{\vartheta}{2} \sim \gamma\vartheta$ with $\vartheta$ the photon emission angle. The figure clearly shows that the low-$s$ part of the full QED spectrum corresponds to wide-angle photon emission, as the only spectral contribution at small $s$ comes from the spectral line characterised by $s \simeq 2b/r_\perp^2$, which is in fact the \emph{linear} Compton line. This relation between small-$s$ and large angles raises an important issue regarding numerical implementation of emission rates. MC codes typically assume photon emission parallel to the electron momentum direction, but we can now see that this is not applicable at small $s$, where photons should rather be emitted at wide angles.  

\begin{figure}[t!]
	\begin{center}
	\includegraphics[width=0.99\columnwidth]{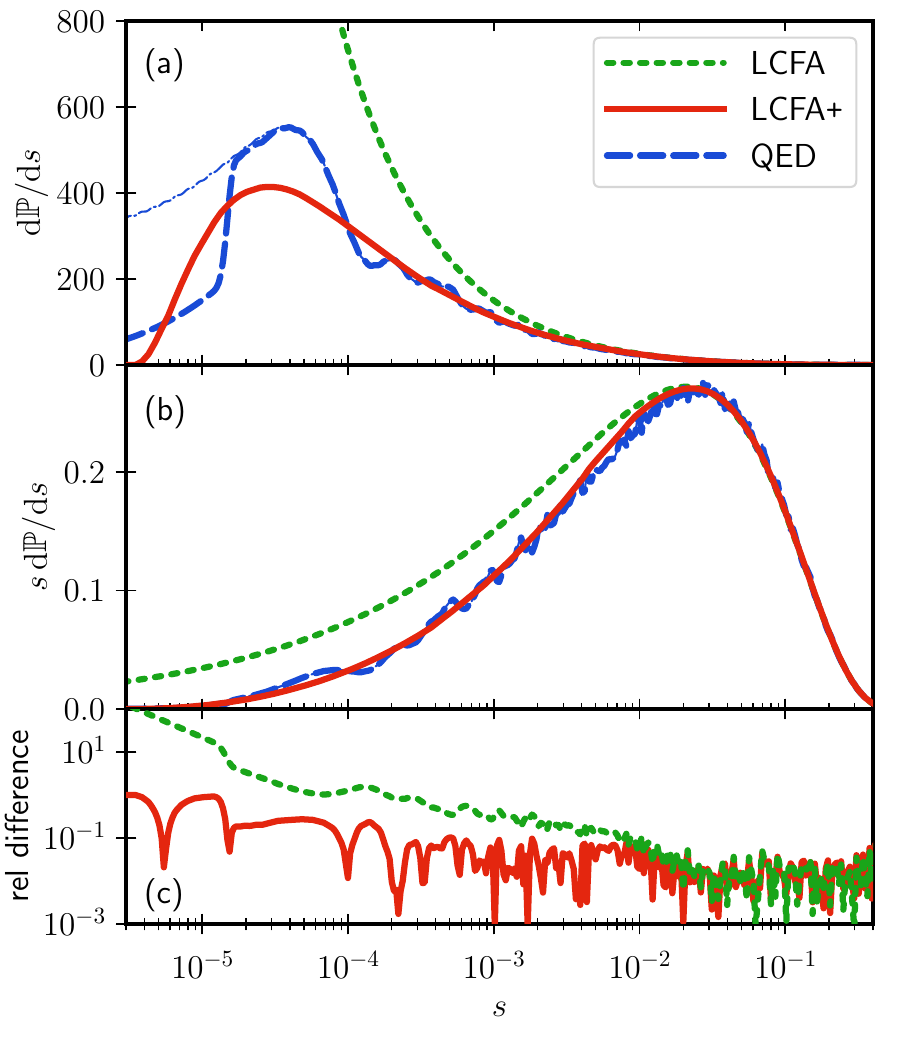}
	\includegraphics[width=0.99\columnwidth]{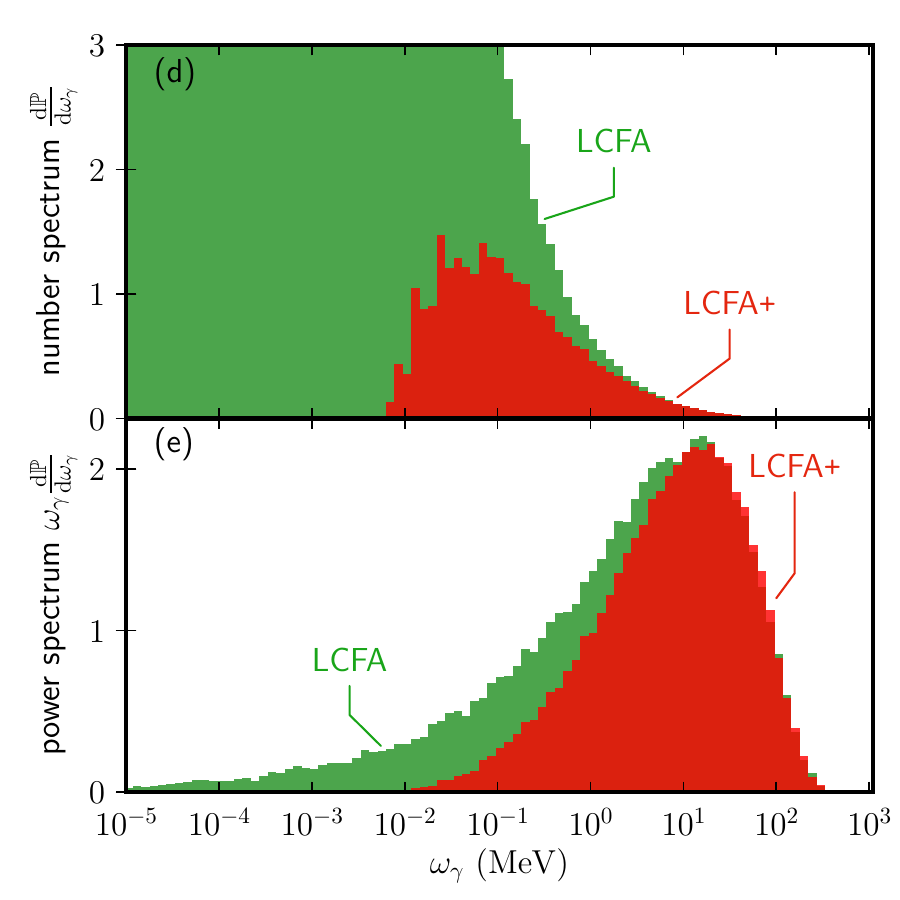}
	\end{center}
	\caption{Comparison of the \lcfap results with QED for $a_0 = 10$, $\tau=4$, linear polarisation and $\gamma=2000$,  (a--c). Thick blue dashed curves labelled QED include a finite angle cutoff, while the thin curves do not. The \lcfap shows improvements over the whole spectrum, much more closely tracking the QED result  than the LCFA does. Panels (d) and (e) show the photon spectra, as a function of emited frequency, from a Monte Carlo simulation at $a_0=7$ and  $\tau=23$.}
	\label{fig:lcfa+pulse}
\end{figure}

Because of this, and because the LCFA fails at low-$s$, we conclude that it is advisable to \emph{exclude} the low-$s$ part of the photon spectrum in MC codes. Indeed, a low-energy or low-$s$ cutoff \footnote{
The specific value depends on the physical scenario to be studied. For instance, a cutoff on the order of the electron mass has been suggested for the investigation of cascades~\cite{elkina11}, while a cutoff in $\chi_\gamma = \mathcal O(10^{-5})$ has been used in single-particle simulations of radiation reaction~\cite{Green:2013sla,harvey15b}.} is often implemented in simulations to prevent the emission of large numbers of low-energy photons originating in the infrared divergence of the LCFA rates~\cite{elkina11,Gonoskov:2014mda}.
 We therefore choose to impose the required positivity condition by removing all contributions at small-$s$, both from the LCFA and its corrections. We do so by imposing a ``positivity filter'', multiplying the intensity-filtered rate by a Heaviside function of the form $\mathbf{\Theta}_{p} :=\Theta(\ud \mathbb{R}_\mathrm{LCFA}/\ud s + \mathrm d\, \delta \mathbb{R}/\ud s )$. Altogether, we define the \lcfap rate as
\be\label{eq:LCFA+}
  \frac{ \ud \mathbb R_\mathrm{\lcfap} }{\ud s} 
 \equiv
 \bigg(  \frac{ \ud \mathbb R_\text{LCFA}}{\ud s}
 +  \frac{ \ud \delta \mathbb R}{\ud s}\,\mathbf \Theta_{i}  \bigg)
\mathbf \Theta_p \;.
\ee
This is positive and well-behaved by construction, and we will now show that \eqnref{eq:LCFA+} approximates QED results to a better degree than the LCFA.  In Fig.~\ref{fig:lcfa+pulse} (a)--(c) (left hand panels) we consider a short laser pulse with envelope $g=\cos^2(\varphi/4\tau) \Theta(2\pi\tau - |\varphi|)$. The introduction of the positivity filter means we are not just adding a corrrection to the LCFA, but we are redefining the LCFA in an improved form. We plot the QED emitted photon number spectrum $\ud \mathbb P /\ud s$ and spectral energy density $s \ud \mathbb P /\ud s$, along with the same spectra calculated with an angular restriction on the emitted photon momentum. In the appendix we derive analytically the appropriate cutoff angle for the tractable cases of linear and circular polarisation. Here we take the intermediate value $r_\perp < 7a_0$.  The only difference in the spectra is at small $s$, for the reasons given above. We also plot the LCFA, and our \lcfap. For large $s$, above the angular cutoff, the \lcfap shows a significant improvement over the LCFA. This is particularly clear in Fig.~\ref{fig:lcfa+pulse} (b), which shows that the LCFA over-estimates the emitted energy, whereas the \lcfap does much better. Turning to small~$s$, we see that the behaviour of the \lcfap (in which emission at low $s$ is removed by the positivity filter) matches with that of the angularly restricted QED rate. Numerical testing shows that the results are insensitive to the precise value of the intensity cutoff $c$ for $1\lesssim c \lesssim 2$ for all $a_0\geq 5$; in these examples we took $c=\pi/2$. A series of further examples are provided in the appendix, all showing improvement over the whole emission spectrum for a wide range of parameters including, notably, intensities as low as $a_0=2$~\cite{LUXE}.

Having now formed an \lcfap which is a demonstrable improvement on the LCFA,  we turn to numerical implementation. In MC-based codes particles propagate (according to the Lorentz force equation) over discrete time steps between instantaneous quantum emission events~\cite{Gonoskov:2014mda}. Such codes allow us to model multi-stage photon emission and pair creation processes which cannot be calculated analytically~\cite{nerush11,king16}. Ideally, we would like to simply replace the LCFA rate in existing codes with the \lcfap rate. However, our considerations so far have been for plane waves, where laser phase $\varphi$ is the natural evolution parameter~\cite{Neville:1971uc,Bakker:2013cea}, and where only phase derivatives of the laser field can occur. For use in simulations  we need to extend our results to more realistic laser fields. First, we extend the variable $\chi_e$ to arbitrary fields using, as in existing approaches, its universal definition $\chi_e = (e/m^2)\sqrt{u.F^2.u}$ in which $u$ is the instantaneous classical four-velocity of the particle between emission events. Second, we convert from $\ud \mathbb P/\ud \varphi$, the probability rate per unit phase to a rate per unit time $\ud \mathbb{P}/\ud t$. This replaces the prefactor $\alpha/b$ with $m \alpha /\gamma(t)$~\cite{ritus85}. Next we turn to the $\mathcal{F}_j$, which at first sight seem intrinsically tied to plane waves.  Applying the Frenet-Serret formalism~\cite{Formiga2006} we find that the $\mathcal F_j$ containing the field derivatives can be written in terms of proper-time derivatives of the four-velocity as 
\begin{align}
\frac{\mathcal F_1}{a_0^2} = - \frac{(\ddot u. \ddot u) +3 (\dot u. \dddot u)}{45(\dot u.\dot u)^2} \,, \quad
\frac{\mathcal F_2}{a_0^2} =    \frac{4 (\ddot u. \ddot u) - 3 (\dot u. \dddot u)}{45(\dot u.\dot u)^2} \;. \label{eqn:F2g}
\end{align}
The right-hand-sides of (\ref{eqn:F2g}) make no explicit reference to the field in which the particle moves, and therefore generalise the plane wave $\mathcal{F}_j$ to arbitrary fields. They can be determined from simulated particle trajectories. (Encouragingly, ${\mathcal F}_1$ also appears in corrections to synchrotron motion, i.e.~non-plane-wave fields~\cite{Khokonov:PRL2002}.) Keeping in mind that current codes assume a high-energy approximation, one may simply take the dots in (\ref{eqn:F2g}) to be time derivatives as a first approximation. Generally, the proper-time derivatives may be traded for time derivatives using $\dot {f} = \gamma(t)\ud f /\ud t$.

Finally, we need a notion of intensity, not dissimilar to that in a plane wave, in order to generalise the intensity filter.  This is a potential restriction on using the \lcfap. 
Now, at high energy, as assumed in current codes, particles see any field as effectively plane wave in a head-on collision~\cite{ritus85,baier-ultra-89,dipiazza13b}. Using this, the transverse kick of an electron (relative to its direction of motion) across a simulation timestep, divided by the electron mass, gives the needed measure of the intensity, as for plane waves. One can be more explicit for a primary case of interest, namely focussed laser beams, where there is a natural laser direction and central frequency. This defines a laser momentum $k_\mu$ so intensity can be defined by $a_0 = m\chi/(k.u)$ as for plane waves~\cite{ilderton09}. This completes our candidate general \lcfap prescription.

As a first test we have implemented the \lcfap in a single-particle code~\cite{harvey15b}. In Fig.~\ref{fig:lcfa+pulse} (d--e) we show the results of an experimentally relevant simulation of a $1$ GeV electron beam colliding with a (plane-wave) background laser pulse of $a_0=7$ and $\tau=23$ (duration $45$ fs)~\cite{cole18,LUXE}. The \lcfap results, for which the average number of emission events per simulation run was $n\simeq 5$, follow a similar pattern to the one-photon emission results above, correcting for the overestimate of the LCFA.  We have thus demonstrated that our results can be employed numerically, in the same way as the LCFA, to study multiphoton processes in laser-particle interactions. (Numerical testing in full PIC simulations is underway.)

In conclusion, we have considered corrections to the locally constant field approximation, LCFA, of nonlinear Compton scattering. As presented in the literature (see e.g.~\cite{baier-mono-81} for the monochromatic case), these corrections are not well behaved. The LCFA is, though, the first term in an asymptotic expansion of the QED result, and thus corrections should be treated as appropriate for an asymptotic expansion. If the asymptotic parameter is not large, then these corrections should not be included. Physically, the reason for this difficulty is that while the LCFA is intended to work at large $a_0$, this must be understood as a local statement, and in a pulse $a_0(\varphi)$ cannot remain large indefinitely.

We have shown that by adding the lowest order correction to the LCFA when it is, in a controlled manner, small, and neglecting the correction when it is large (consistent with an asymptotic treatment), that we can generate a positive, well-behaved rate which gives a significantly improved approximation to the full QED result in plane wave backgrounds.  The neglect of the LCFA corrections is also physically motivated.  Our results hold over a range of intensity and energy parameters relevant to current and upcoming laser experiments. We have also demonstrated the numerical implementation of our results in a single-particle Monte-Carlo code. Although we focussed on nonlinear Compton scattering, but our calculations can be extended directly to the process of nonlinear Breit-Wheeler, the second quantum process usually included in simulations.

\textit{The authors thank T.~Blackburn, A.~Di~Piazza and A.~G.~R.~Thomas for useful discussions. The authors are supported by the EPSRC, grants EP/S010319/1 	 (AI \& BK) and by the STFC, grant ST/G008248/1 (DS). BK acknowledges the direct support of EP/P005217/1 for this work.} \\[-10pt]

\noindent{\href{mailto:anton.ilderton@plymouth.ac.uk}{\scriptsize anton.ilderton@plymouth.ac.uk}} \\[-2pt]
{\href{mailto:b.king@plymouth.ac.uk}{\scriptsize b.king@plymouth.ac.uk}} \\[-2pt]
{\href{mailto:dseipt@umich.edu}{\scriptsize dseipt@umich.edu}}
%
%

%




\appendix


\section{Derivation of the LCFA and its corrections}

Let $k_\mu$ be a null vector, so $k^2=0$, defining the propagation direction of the plane wave. We can always take $k.x = \omega(t+z)$, lightfront time, where $\omega$ is e.g.~the central frequency, used to define dimensionless variables. The plane wave is then described by a potential $e A_\mu(x) = m a_\mu(k.x)$ with only nonzero ``transverse'' components $\vec a_\LCperp = \{a_x, a_y\}$, see e.g.~\cite{dinu12,Seipt:2017ckc}. The dimensionless electric field variables used in the text, $\vec \varepsilon_\LCperp(k.x) \equiv e \vec E_\LCperp(k.x) /m\omega$, are then related to the potential by $\vec \varepsilon_\LCperp(k.x) = \vec a'_\LCperp(k.x)$. Recall that we decompose the field components into peak amplitude $a_0$ and profile functions $h_j$ by writing $\varepsilon_j(k.x) = a_0 h_j(k.x)$.

The probability of photon emission in the plane wave is an integral over $s$ (the emitted photon momentum fraction introduced in the text) and two lightfront times, or phases, $\varphi$ and $\theta$ arising as the average and difference of the interaction point phase in the scattering amplitude and its complex conjugate. As a result, $\theta$ is naturally associated with quantum interference effects, see~\cite{dinu16}. In terms of $\varphi$ and $\theta$, we define the floating average $\langle\cdot \rangle$ over the phase interval $\theta$ by
\be
	\langle f \rangle = \frac{1}{\theta} \int\limits_{\varphi-\theta/2}^{\varphi+\theta/2} \!\ud (k.x)\, f(k.x) \;,
\ee
and from this Kibble's (normalised) effective mass~\cite{Kibble:NPB1975,harvey12}
\begin{align}
\mu(\varphi,\theta)= 1 + \langle \vec a_\LCperp^2 \rangle - \langle \vec a_\LCperp\rangle^2 \,.
\end{align}
In terms of $\mu$, the energy parameter $b$, and the photon momentum fraction $s$, the total emission probability is compactly expressed as~\cite{dinu13a}
\begin{align}\label{P-starting-point}
\mathbb{P}	= -\frac{\alpha}{\pi b} \int\limits_0^1\!\ud s \!\int\! \ud\varphi\! \int\limits_0^\infty\!\ud\theta \: \sin ( x_0 \theta\mu )
	\bigg[ \frac{1}{\mu}\frac{\partial\mu}{\partial\theta} +  g \langle   a' \rangle^2 \theta \bigg] ,
\end{align}
where the leading $1/b$ comes from state normalisation~\cite{Ilderton:2012qe}, $x_0 := s/2b (1-s)$, the function $g$ contains spin effects,
\be\label{g-def}
	 g := \frac{1}{2} + \frac{1}{4}\frac{s^2}{1-s} \;,
\ee
and the integrand of (\ref{P-starting-point}) is a function of $a_0$ and $b$, in general, not of $\chi$.

The LCFA is usually said to hold at $a_0\gg 1$, and has been obtained in various cases as an asymptotic limit, in which emission probabilities are functions of $\chi$ alone, up to normalisation. It is however not immediately obvious how to include corrections to the LCFA starting from the general QED expressions above, due to the complexity of the multi-dimensional integrals which must be performed. Existing literature suggests though that the LCFA is related to a \textit{small~$\theta$} expansion of the probability~\cite{Khokonov:PRL2002,harvey15,Blackburn:2018sfn}.  We will use this to express the desired asymptotic expansion to a perturbative expansion in powers of $1/a_0$, in doing so encountering a condition which indicates when the expansion breaks down.

\begin{figure}[t!]
	\includegraphics[width=\columnwidth]{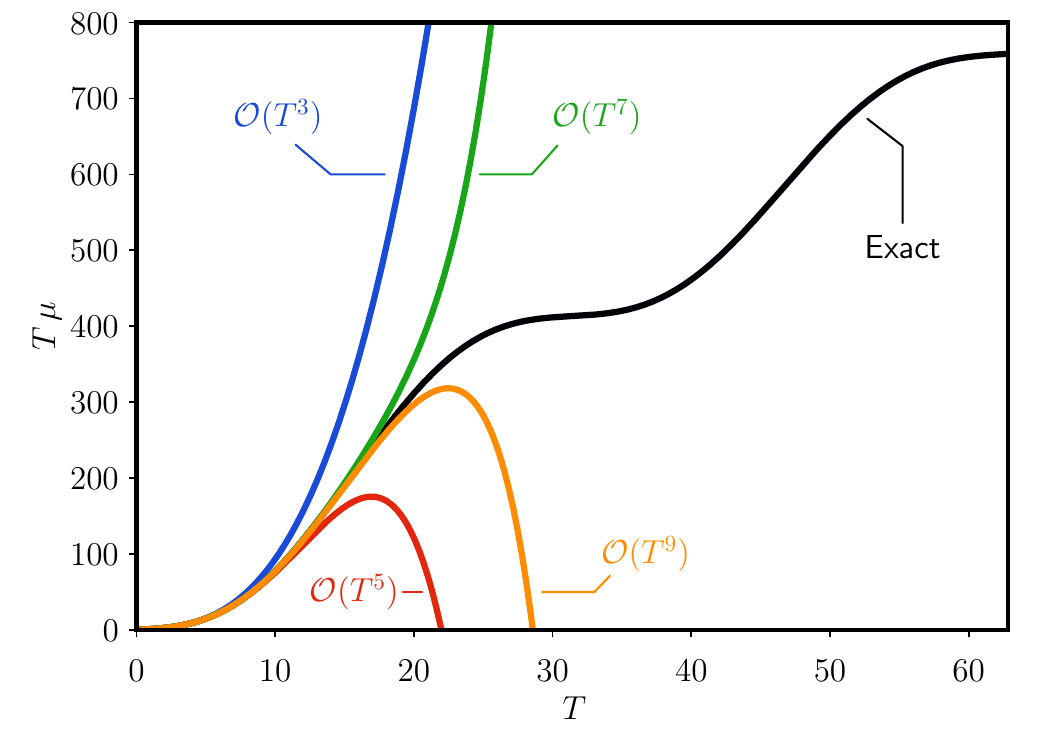}
	\caption{\label{FIG:M2-EXPANSION} The expansion of $T$ times the effective mass, $T \mu(\varphi, T/a_0)$ for a linearly polarised Gaussian pulse with $h = \sin(\phi)\exp(-\phi^2/\Delta^2)$. The expansion to several orders $T^n$ is shown ($\varphi=0$, $a_0=5$, $\Delta=10$) for $n$ from 3 (giving the LFCA) to $9$. As $n$ increases the small $T$ behaviour improves, but the large $T$ behaviour worsens, leading to problems at small~$s$ as discussed in the main text.}
\end{figure}

First, rescale $\theta$ to a new variable $T =  a_0 \theta$. Doing so turns the integrand into a function of $a_0$ and $\chi$; observe that the argument of $\sin(\cdot)$, which is the only place $b$ appears, behaves as
\be\label{A4}
	x_0 \theta \mu(\varphi,\theta) \to
	\frac{s}{2\chi(1-s)} {T} \mu(\varphi, {T}/a_0) \;.
\ee
We then expand the entire integrand in powers of $1/a_0$, at fixed $\chi$.  The lowest order terms are \textit{independent} of $a_0$ (and correspond to the formal limit $a_0\to\infty$ at fixed $\chi$).  Using (\ref{A4}) to illustrate this point we find
\be \label{eq:phase-cubic}
	x_0 \theta \mu(\varphi,\theta) \approx	\frac{s}{2\chi(1-s)} \bigg( {T} + \frac{1}{12} {T}^3 h(\varphi)^2 + \mathcal{O}(a_0^{-2}) \bigg)\;.
\ee
Fig.~\ref{FIG:M2-EXPANSION} shows different orders of this expansion. The key point is that this rescaling and expansion turns the Kibble mass into, at lowest order, a cubic function, which is typical of the constant crossed field case. Indeed these terms lead (see immediately below) to the LCFA. The higher order terms, which begin with a power of $1/a_0^2$, are to be expanded out, and hence seem to give corrections to the LCFA in powers of a small parameter. (See also~\cite{Dinu:2017uoj} for a $1/a_0$ expansion of the trident process.) 

To be explicit, consider first only the lowest order terms. However, to obtain the LCFA we need to perform a further change of variables from ${T}$ to a new variable, $t$, such that the argument of $\sin(\cdot)$ is brought into Airy form proper. The required change of variable is
\be
	{T} = \bigg(\frac{8\chi(1-s)}{h^2(\varphi)s }\bigg)^{1/3} t \;.
\ee
Proceeding from here to evaluate the $t$ integrals yields the LCFA to zeroth order, and then the corrections. However, before describing the calculation, we emphasise an important point. We began with the QED probability and made an expansion in $1/a_0$, small. In order to bring the integrals to the required form to reproduce the LCFA and LCFA\texttt{+}, though, means using an overall change of variables
\be\label{A8}
	\theta =  \bigg(\frac{8b(1-s)}{a_0^2(\varphi)s }\bigg)^{1/3} t \;,
\ee
and an expansion in powers of $\theta$. Hence the change of variables needed to pass from the general QED result (\ref{P-starting-point}) to the known LCFA (plus corrections) is singular when \textit{local} $a_0(\varphi)\to 0$.  Considered as a \textit{perturbative series}, the expansion \eqref{eq:phase-cubic} represents a good approximation when higher-order terms that are present, but not included, are negligible. In terms of the old and new variables, this condition corresponds to the coefficient relating $\theta$ to $t$ in (\ref{A8}) remaining small, as otherwise higher powers in the Taylor series will dominate lower powers. This holds only when
\be\label{exp-p}
	\zeta^{-1/3} := \bigg(\frac{8b(1-s)}{a_0^2(\varphi)s }\bigg)^{1/3} \ll 1 \;,
\ee
which is violated when $a_{0}(\varphi) \to 0$, \emph{locally}, or when $s\to 0$. This confirms earlier results~\cite{khok,dinu16} and provides a straightforward derivation of the result that (\ref{exp-p}) is a relevant expansion parameter for the LCFA~\cite{meuren17}. We interpret this breakdown at small $a_0(\varphi)$ in the sense of an \textit{asymptotic} series as giving a condition for when corrections to the asymptotic result (the LCFA) should be included or discarded. The physical justification is clear; emission rates are small in regions of low laser-intensity, and already the LCFA rates are small there, producing the expected physics without need of a correction. The mathematical justification is that in regions of low laser-intensity, the effective asymptotic parameter $\zeta$ is no longer large and asymptotic corrections are therefore not accurate.
 
Proceeding, the leading order term of our expansion brings the probability to the form
\be\begin{split}\label{Dave}
	\mathbb{P} \simeq -\frac{\alpha}{\pi b}\int\!&\ud\varphi\!\int\limits_0^1\!\ud s\!\int\limits_0^\infty\!\ud t\ \sin\big(zt + \tfrac{1}{3}t^3\big) \bigg[  \frac{2t}{3z+t^2} -  \frac{4t}{z}  g\bigg] , 
\end{split}
\ee
in which $z$ is given by (\ref{eqn:zg}) in the text. It remains only to perform the $t$ integrals, turning them into the Airy functions familiar from the constant field case. The term containing $g$ is simplest:
\be\label{Colin}
	\int\limits_0^\infty\!\frac{\ud t}{\pi}\ t \sin\big(zt + \tfrac{1}{3}t^3\big)=-\text{Ai}'(z) \;.
\ee
Turning to the first term in square brackets of (\ref{Dave}), we introduce a parameter integral to write the integrand in terms of cosine, then perform the $t$ integral to obtain the second derivative of the Airy function; using Airy differential equation one then obtains
\be\begin{split}\label{Kate-2}
	&-\frac{2}{3}\int\limits_0^\infty\!\frac{\ud t}{\pi}\int\limits_1^\infty\!\ud\alpha \ t^2 \cos{\left[\alpha\big( zt + \tfrac{1}{3}t^3\big)\right]}	 \\
	&= \frac{2}{3}\int\limits_1^\infty\!\ud\alpha \ z \alpha^{-1/3} \text{Ai}( z \alpha^{2/3}\big) \;.
\end{split}
\ee
To bring the integral into a more standard form we change the integration variable to $\beta$ defined by $z\alpha^{2/3}= z + \beta$, giving
\be\label{Kate-4}
	\mathrm{(\ref{Kate-2})} = \int\limits_0^\infty\!\ud\beta\ \text{Ai}( z+\beta)  \equiv \text{Ai}_1(z)\;.
\ee
Thus we have
\be
	\mathbb{P} \simeq -\frac{\alpha}{b}\int\!\ud\varphi\!  \int\limits_0^1\! \ud s\,  \text{Ai}_1(z) + \bigg(\frac{2}{z}+\chi_\gamma\sqrt{z}\bigg)\text{Ai}'(z) \;,
\ee
which is precisely the LCFA approximation to NLC. It is common in the literature to replace the $s$-integral with an integral over local $\chi_\gamma(\varphi)=a_0(\varphi)bs$, for which 
\be
	\int\limits_0^1\!\ud s = \int\limits_0^{\chi_e(\varphi)}\! \frac{\ud \chi_\gamma(\varphi)}{\chi_e(\varphi)} \;.
\ee
This completes the calculation of the LCFA terms. The first correction to the LCFA, (\ref{delta-LCFA}) in the main text, is found by including, in (\ref{P-starting-point}) and (\ref{A4}), terms of order $1/a_0^2$, expanded perturbatively. The $\mathcal O(a_0^{-2})$ term in \eqref{eq:phase-cubic} is, for example, see also Fig.~\ref{FIG:M2-EXPANSION},
\begin{align}
\frac{{T}^5}{a_0^2} \frac{h_j'h_j' + 3h_jh_j''}{720} \,.
\end{align}
When expanded out this gives $\mathcal{F}_1$ multiplying the same trigonometric/exponential functions as appear in the LCFA terms, which again yield Airy functions of the same argument. Similarly, the expansion of the exponential and of the average appearing outside it in (\ref{P-starting-point}) generates $\mathcal{F}_2$. The explicit calculation of these terms proceeds similarly to that for the LCFA.

\section{Examples of the improved LCFA}
 
In this section we provide a series of examples illustrating our improvement of the LCFA over a wide range of parameters corresponding to $\chi_e = 0.024 \ldots 0.91$. We compare with both the full QED rate and the angularly restricted QED rate, the latter comparison serving to illustrate that the effect of our filters is essentially the same as imposing an angular cutoff on the QED rates.

\begin{figure*}[t!]
	\begin{center}
		\includegraphics[width=0.45\textwidth]{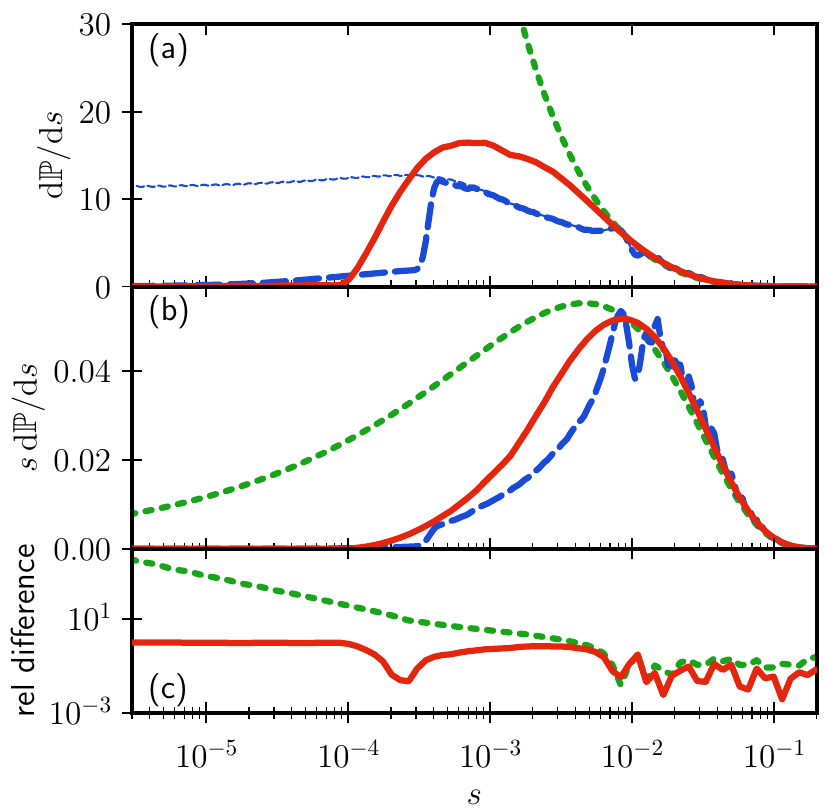}
		\includegraphics[width=0.45\textwidth]{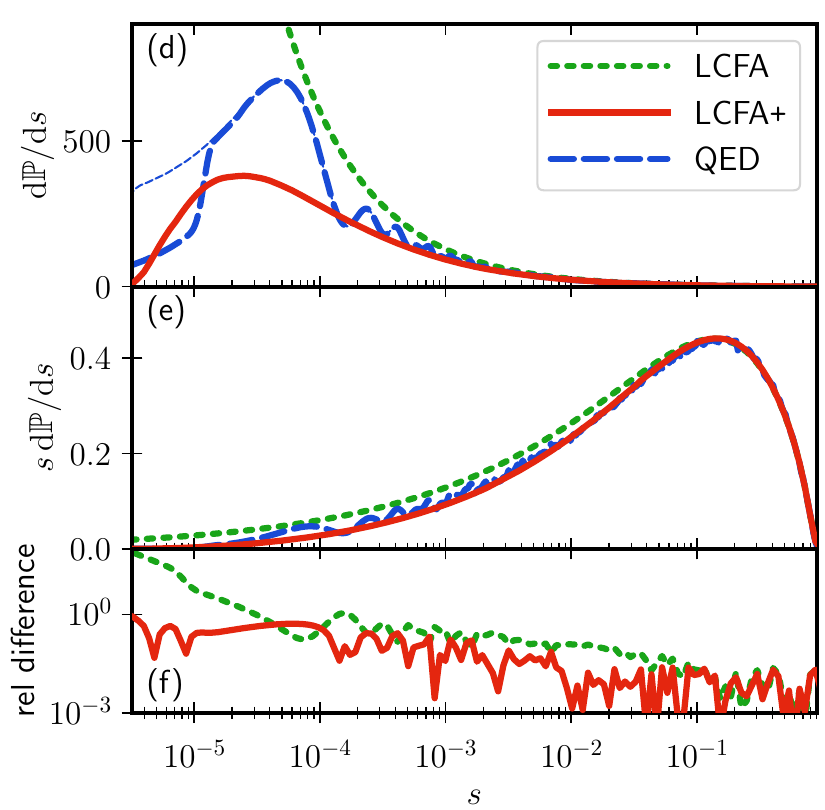}
		\includegraphics[width=0.45\textwidth]{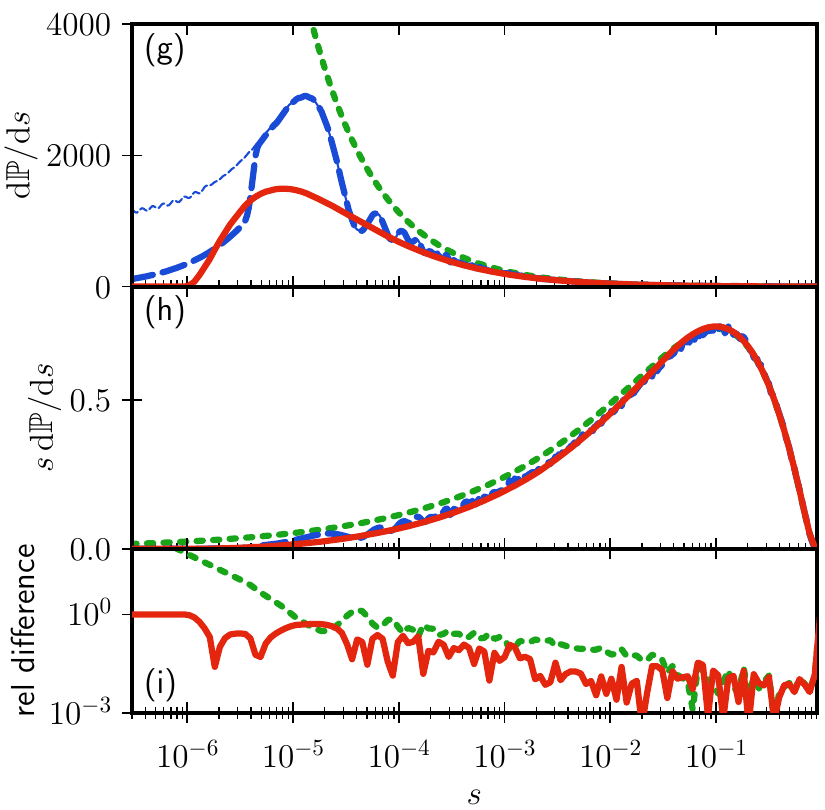}
		\includegraphics[width=0.45\textwidth]{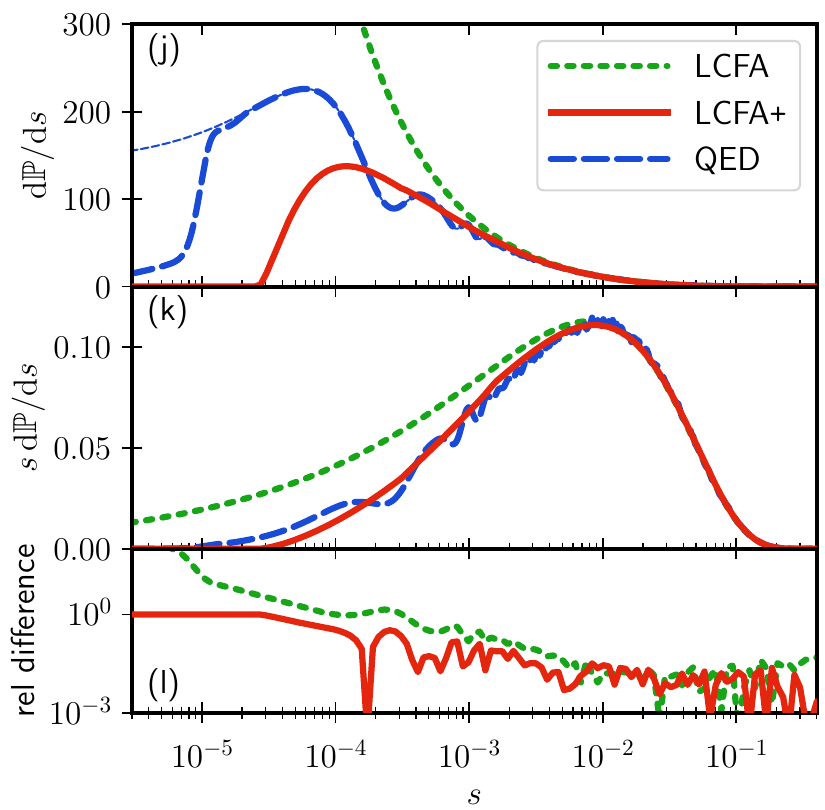}
	\end{center}
	\caption{Comparison of the \lcfap with the QED result (blue dashed, thin lines), and the angularly restricted QED result (blue dashed, thick lines), for linear laser polarisation and
		(a--c): $a_0 = 2$, $\gamma=2000$,
		(d--f): $a_0 = 25$, $\gamma=10000$,
		(g--i): $a_0 = 30$, $\gamma=5000$;
		{and for circular polarization with $a_0=5$, $\tau=2$ and $\gamma = 1000$ (j--l).} 
		There is improved agreement compared with the LCFA, and the effect of the filters is clearly comparable with the imposition of an angular cutoff. Note that for (a--c) a lower intensity filter cutoff of $c=0.5$ was used since the usual value is too close to the peak $a_0$. The finite-angle condition imposed was, in all cases,  $r_\perp < 7a_0$.
}
	\label{fig:comp-app}
\end{figure*}

Fig.~\ref{fig:comp-app} shows the improvement of the \lcfap relative to the normal LCFA, even in the low $s$ region where the improved rate matches much better to the angularly resolved QED rate. It is remarkable that the \lcfap works well even down to $a_0 = 2$, where one would not expect local approximations to hold. This is also relevant for future laser experiments~\cite{LUXE}.

\begin{figure}[t!]
	\begin{center}
		\includegraphics[width=0.99\columnwidth]{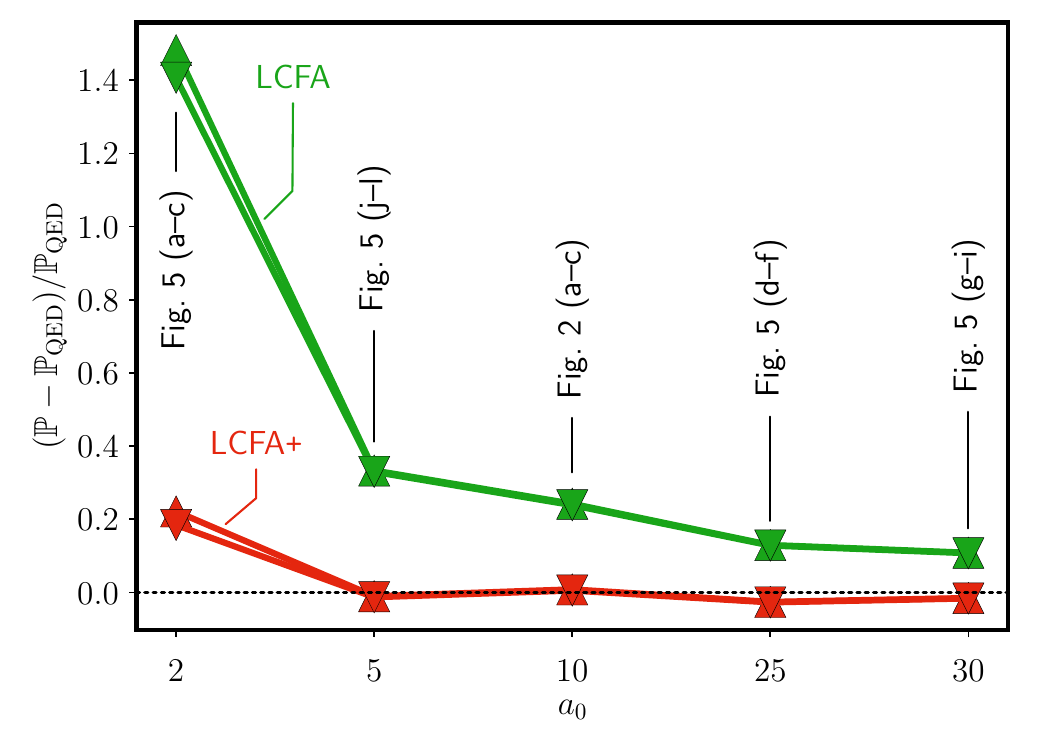}
	\end{center}
	\caption{Quantification of the improvement of \lcfap (red) compared to LCFA (green) for various values of $a_0$ (not a linear scale). Both the full and angularly restricted QED probability are used
		(symbols $\bigtriangleup$ and $\bigtriangledown$, respectively).}
	\label{fig:quantitative}
\end{figure}

A quantitative analysis of the improvements is given in Fig.~\ref{fig:quantitative}, where we list the relative differences of the total probabilities as given by the \lcfap and the LCFA for all the examples here and in the main text. This confirms that the improvement of the \lcfap over the LCFA is significant. (We find that it is unimportant, for this comparison, whether we compare with the angularly restricted or full QED probability).

\section{Estimation of the size of the LCFA and LCFA\texttt{+}}

As already explained, the global positivity filter in the \lcfap implies that it cannot be reduced to a form ``LCFA plus correction term''. In this section we use the word ``correction'' to refer to the extra terms introduced by the derivative expansion.
Here we estimate the size of the different terms of the LCFA, Eq.~\eqref{Dave-4}, as the leading order asymptotic approximation of the  full QED expression, and the higher-order derivative terms, Eq.~\eqref{delta-LCFA}, as next-to-leading-order terms. 
For $a_0^{2}/b$ large and $s$ not too small the asymptotic parameter $\zeta=a_0^2 s/ 8b(1-s)$ is large~\cite{meuren17} and the LCFA by itself represents a good approximation, with the correction terms small and well-behaved.

Looking instead at the infrared behaviour for small $s\ll1,\chi_e$ we find three different classes of terms. 
These are IR finite terms, integrable IR-divergent terms $\propto s^{-2/3}$ occurring both in the LCFA and the higher-order derivative terms, and a non-integrable IR-divergent term $\propto s^{-4/3}$ in the correction. We only need to discuss the latter two cases, starting with the integrable terms.

The IR-divergent, but integrable, term of the LCFA behaves as $ \mathrm{Ai}'(0) ({s}/{\chi})^{-2/3}$, and the corresponding terms in the correction are 
\be
	\frac{\mathcal F_2}{a_0^2} \mathrm{Ai}'(0) \left(\frac{s}{\chi}\right)^{-2/3}\quad \text{and} \quad \frac{\mathcal F_1}{a_0^2} \mathrm{Ai}'(0) \left(\frac{s}{\chi}\right)^{-2/3}.
\ee
This means that the correction terms remain small when
${\mathcal F_j}/{a_0^2} \ll1$. Hence, as discussed above, the intensity filter excludes the corrections when the local value of $a_0$ is small and the corrections become large.  In principle this also sets limitations on the field gradients (which are hidden in the $\mathcal F_j$) which is not surprising given that the the corrections have been expressed as a gradient expansion.
Taking for the sake of definiteness the case of a circularly polarized plane wave, where $\mathcal F_1 = -2/45$ and $\mathcal F_2= -7/45$, we can conclude that even for intensity filter cut-off values $O(1)$ the IR-finite corrections are well-behaved.

Now consider the non-integrable IR divergent term in the \lcfap. As $s\to 0$, this term $\propto s^{-4/3} $ will outgrow the LCFA IR-divergent term going like $s^{-2/3}$. The fact that already the LCFA is IR-divergent, in contradiction to the exact QED result which approaches a finite value, has prompted attempts to fix the LCFA \cite{meuren17}, or more often to just impose a low-energy (or small-$s$) cutoff.
In our analysis, however, we now also have the next-to-leading order correction and we can compare the two for small~$s$, giving further insights on the accuracy of the asymptotic expansion. The point $s=s_\star$ where the correction is the same size as the leading order term can be calculated as
\begin{align} \label{eq:sstar}
s_\star & = \frac{2 b}{a_0^2(\varphi) } \frac{||\mathcal F_2||^{3/2}}{2} 
\left( \frac{\mathrm{Ai}(0)}{ \mathrm{Ai}'(0)}\right)^{3/2} \,.
\end{align}
Then, according to the IR-behaviour of the two terms we find the approximate scaling
\begin{align}
\frac{\delta \mathbb R}{\mathbb R_\mathrm{LCFA} }
 \sim \left( \frac{s}{s_\star} \right)^{-2/3} \,.
\end{align}
That means one order of magnitude above $s_\star$ the ratio is only $0.2$, 2 orders of magnitude above it is only $0.05$ and 3 orders of magnitude above $s_\star$ it is only $1\%$. This analysis only involves the leading IR divergent terms and is strictly only valid for small $s \ll1$. Taking again the case of circular polarization to estimate the value of the field gradients in $\mathcal F_2$ we find
$s_\star  \approx 0.05  \frac{2 b}{a_0^2(\varphi) } $.
For linear polarization, we can find the approximate values
for $s_\star \approx 0.013 (2 b/a_0^2(\varphi)) $ for laser phases close to the peak of the electric field.

In the \lcfap we add the LCFA and the correction. Because $\mathcal F_2$ is negative the \lcfap rate turns negative for 
$s<s_\star$, and these negative values are removed by the positivity condition for the rate discussed in the main text. 

How does $s_\star$ compare to typical frequencies of the emitted photons? To answer this we compare with the first nonlinear Compton edge, i.e.~with~the smallest on-axis ($r_\perp=0$) frequency of the red-shifted first non-linear Compton harmonic, which is characterized by, for circular polarisation,
\begin{align}
\label{eq:NLKN}
s_\mathrm{1} = \frac{2b}{1 + 2b + r_\perp^2 + a_0^2 } \,.
\end{align}
Assuming $b \ll 1 + a_0^2$ and setting $r_\perp=0$ we find that $s_\star < s_\mathrm{1}$ for $a_0 \gtrsim 0.23$ and 
$s_\star < 0.1 s_\mathrm{1}$ for $a_0 \gtrsim 1$. This means, for all relevant cases, the radiation emitted with  light-front momentum fraction $s_\star$ (where the rate would become negative without the positivity condition) is emitted well below the first Klein Nishina edge. Such radiation must therefore be emitted under a large angle $r_\perp \gg 1$. 
Solving \eqref{eq:NLKN} with $s_\mathrm{1} =s_\star$ we find that
the typical angle $r^\perp_\star \approx \gamma\theta_\star \approx 4.35 a_0$.
Repeating the calculation for linear laser polarisation, we find instead $r^\perp_\star \approx 8.4 a_0$. In both cases photons are emitted far outside the usual $1/\gamma$ radiation cone. These results motivate our choice for the angular restriction of the QED emission used in the text, taking a mid-point value $r_\perp < 7 a_0$ for comparison with the \lcfap.



	We consider a numerical example, the collision of a 10 GeV electron beam with a laser pulse of $a_0=5$ and $\omega= 1.55$ eV. This gives $b= 0.12$. This corresponds to a cutoff value of $s_\star = 4.8 \times10^{-4}$ which corresponds to a photon energy of the order of $\omega_\star = 4.8$ MeV. According to the prescription above all photons with energy below $\omega_\star$ should be considered as ``low-energy'' and discarded from the simulation. How can it be justified that photons with energies higher than the electron rest mass should be neglected? The answer lies in two points.
		
	First, $s_\star$ or $\omega_\star$ should be compared to the typical frequencies in the photon spectrum.
	For instance, we can compare to the Compton edge, which in this case is $s_1=9\times 10^{-3}$ or 
	$\omega_1 = 90$ MeV, and which represents the typical energy in the low-energy part of the QED spectrum. Most of the emitted photons will have energies larger than $\omega_1$. The mean energy of the emitted photons can be estimated using the constant crossed field results as being on the order of $\langle \omega \rangle = 870$ MeV. Since the constant crossed field results are affected by the infrared divergence we expect the full QED result to be even slightly higher \cite{ritus85}.
	The critical energy (which bisects the power spectrum) is estimated as $\omega_\mathrm{crit} = 2.4$ GeV, i.e.~half of the electron energy loss is due to photons emitted with energies higher than 2.4 GeV.
	
%
	
	Second, when it comes to the possibility of subsequent pair production by the emitted photons the decisive parameter is $\chi_\gamma$, which, for a head-on collision is
	$\chi_\gamma \approx \frac{2 \omega \omega'}{m^2} \, a_0 \sim s b a_0$. Pair production only occurs with a high probability
	for $\chi_\gamma \sim 1$ and it is exponentially suppressed for $\chi_\gamma \ll1$. For our numerical example we have
	$\chi_\gamma^\star = 1.4\times 10^{-4}\ll1$. Photons below the low-energy cutoff cannot produce paris efficiently.
	More generally we can estimate that $\chi_\gamma^\star \sim s_\star b a_0 \sim 0.1 b^2/a_0^2 \ll 1 $ for $a_0\gg b$.
	The latter condition requires that we are in the \emph{high-intensity} regime of strong-field QED~\cite{khok,dinu16}, in contrast to the high-energy regime, where the behavior of emission rates is quantitatively different~\cite{dinu16,Podszus:2018hnz,Ilderton:2019kqp}.
	The estimates above show that, as a self-contained description of high-intensity laser-plasma interactions, it is reasonable
	to discard all photons below $s_\star$ as they (i) only marginally affect the energy losses of the electrons and (ii) are very unlikely
	to produce pairs. 

\section{Fringe cases with positive $\mathcal{F}_2$}
The sign of $\mathcal{F}_2$ is important for the low-$s$ behaviour of the corrections to the LCFA rate, determining whether the rate goes to plus or minus infinity. For plane wave pulse shapes typically considered in the literature, we have found that $\mathcal{F}_2(\varphi) <0$ is always fulfilled for all reasonable pulse shapes, as we saw in the case of monochromatic fields. It is nevertheless possible to find pulse shapes for which $\mathcal{F}_2(\varphi) >0$ for some $\varphi$. These examples are, though, somewhat contrived, describing non-standard pulse shapes. As such we consider them to be, at least for the case of plane waves, fringe cases. The situation for general fields is less clear, and will be examined in detail elsewhere.

However, as an initial investigation we have performed simulations of the classical propagation of electron bunches through focused Gaussian laser pulses (focal spot $5\,\mu\trm{m}$, $a_0=10...100$) in order to understand this physically relevant case. These calculations show that $\mathcal F_2 < 0$ holds everywhere in the vicinity of the laser focus where $a_0$ is large. We have found that $\mathcal F_2(\varphi) > 0$ only occurs in regions where $a_0(\varphi) \ll1$, far from the pulse focus, and it is not certain if this genuinely is positivity or an effect due to numerical error. We therefore looked for $\mathcal F_2(\varphi)>0$ in all regions where $a_0(\varphi) > 10^{-6}$, and found no occurrences. This implies that the positive values of $\mathcal F_2(\varphi)$, if they exist somewhere, would in any case be removed by the intensity filter already present in the \lcfap for such Gaussian beams.

\begin{figure}[!b]
	\begin{center}
		\includegraphics[width=0.99\columnwidth]{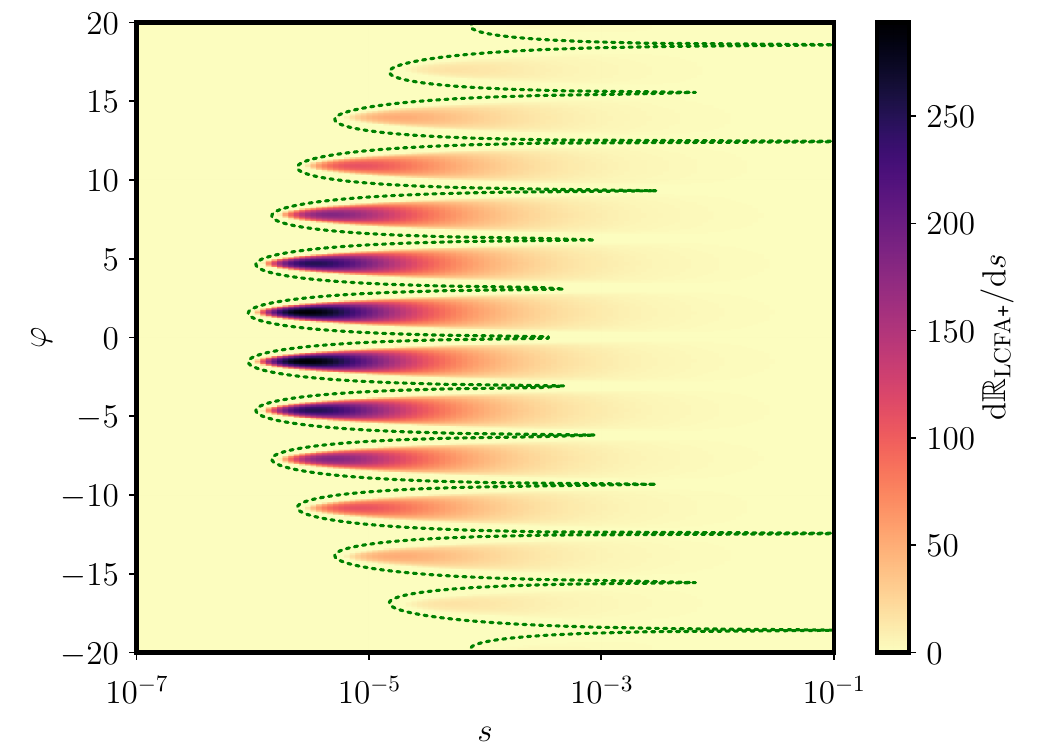}
	\end{center}
	\caption{Contourplot of the \lcfap rate as a function of $s$ and laser phase $\varphi$.
		The green curve is a local low-$s$ cutoff equivalent to a finite emission angle in QED, determined 
		according to $s_\star(\varphi) = 2b/75a_0^2(\varphi)$.}
	\label{fig:map}
\end{figure}

Returning to general fields, the positivity filter discussed in the text protects against the case when $\mathcal{F}_2(\varphi)<0$. In the fringe case when $\mathcal F_2(\varphi)>0$ for some $\varphi$, we can show here that a solution is to impose an additional hard cutoff at small~$s$. To motivate this, consider again Fig. \ref{fig:map}, which shows that the low-$s$ cutoff introduced by the positivity filter matches well with a corresponding large $r_\perp $ cutoff. Indeed, because at large $r_\perp$ only linear Compton scattering contributes to the full rate, see Fig.~\ref{fig:QED-angle}, a small $s$ cutoff can be mapped to a large-angle cutoff as $s>s_\star(\varphi) = 2b/ [ \tilde{c} a_0 (\varphi)]^2$ for some constant  $\tilde{c}$, motivated by \eqref{eq:sstar}. The value of $\tilde c$ can be determined locally from the magnitude of the field gradients.  This means that if we impose a hard cutoff at low-$s$, $s>s_\star(\varphi)$, acting as a failsafe in case $\mathcal F_2(\varphi)>0$, then we can understand the resulting rate simply as being angularly restricted. In Fig.~\ref{fig:map} we used $\tilde{c}^2=75$, effectively multiplying \eqnref{eq:LCFA+} by $\Theta\left(s-s_{\ast}(\varphi
)\right)$. This means that all photons emitted within a cone with aperture angle $\vartheta \simeq 17.3 a_0/\gamma$ are taken into account by the \lcfap rates (and in a MC code they would be emitted parallel to the electron), while photons falling outside this cone are discarded.

%

\providecommand{\noopsort}[1]{}

\end{document}